\documentclass[conference]{iaria}
\usepackage[linesnumbered,ruled,vlined]{algorithm2e}
\usepackage{graphicx}
\usepackage{textcomp}
\usepackage{xcolor}
\usepackage{cprotect}
\usepackage{listings}
\usepackage{subfig}
\usepackage{adjustbox}
\usepackage{tabularx}
\usepackage{hyperref}
\usepackage{makecell}
\usepackage{booktabs}
\usepackage{amsmath}

\AtBeginDocument{
  
}

\title{Addressing Malware Family Concept Drift with Triplet Autoencoder}

\author{
    \IEEEauthorblockN{Numan Halit Guldemir}
    \IEEEauthorblockA{
        \textit{Centre for Secure}     \\
        \textit{Information Technologies}   \\
        \textit{Queen's University Belfast}\\
        United Kingdom \\
        nguldemir01@qub.ac.uk
    }
    
    \and
    
    \IEEEauthorblockN{Oluwafemi Olukoya}
    \IEEEauthorblockA{
        \textit{Centre for Secure}     \\
        \textit{Information Technologies}   \\
        \textit{Queen's University Belfast}\\
        United Kingdom \\
        o.olukoya@qub.ac.uk
    }
    
    \and
    \IEEEauthorblockN{Jesús Martínez-del-Rincón}
    \IEEEauthorblockA{
        \textit{Centre for Secure}     \\
        \textit{Information Technologies}   \\
        \textit{Queen's University Belfast}\\
        United Kingdom \\
        j.martinez-del-rincon@qub.ac.uk
    }
}

\begin{document}
\maketitle

\begin{abstract}
Machine learning is increasingly vital in cybersecurity, especially in malware detection. However, concept drift —where the characteristics of malware change over time— poses a challenge for maintaining the efficacy of these detection systems. Concept drift can occur in two forms: the emergence of entirely new malware families and the evolution of existing ones. This paper proposes an innovative method to address the former, focusing on effectively identifying new malware families. Our approach leverages a supervised autoencoder combined with triplet loss to differentiate between known and new malware families. We create clear and robust clusters that enhance the accuracy and resilience of malware family classification by utilizing this metric learning technique and the Density-Based Spatial Clustering of Applications with Noise (DBSCAN) algorithm. The effectiveness of our method is validated using an Android malware dataset and a Windows Portable Executable (PE) malware dataset, showcasing its capability to sustain model performance within the dynamic landscape of emerging malware threats. Our results demonstrate a significant improvement in detecting new malware families, offering a reliable solution for ongoing cybersecurity challenges.

\end{abstract}


\section{Introduction}

Machine learning has become a key tool in cybersecurity, particularly for detecting malware. These systems, when well-trained, are highly effective at identifying threats. However, the effectiveness of these systems is continually challenged by the dynamic nature of malware. As cyber threats rapidly evolve, machine learning models that were once effective can quickly become obsolete — a phenomenon known as concept drift \cite{lu2018learning}. Concept drift in malware detection is often driven by two major factors: developing entirely new malware and modifying existing malware to evade detection systems. A report by AV-Test indicates approximately 320,000 new malware samples emerge daily, underscoring the need for continuous model adaptation to unseen threats \cite{AVTest}.

Retraining models frequently is one common strategy to fight against concept drift, but it has problems. Firstly, in cybersecurity, accurately labeling data is essential but costly, as it often requires experts to examine and classify new threats \cite{pendlebury2019tesseract}. Moreover, determining the precise timing for updating or retraining a model is not straightforward \cite{jordaney2017transcend}. Frequently retrained models might fail to keep up with the latest malware without a reliable method to decide when updates are necessary, resulting in security gaps.

Labeling unknown samples is crucial to ensure that the model does not misclassify them and can be accurately analyzed by experts. Machine learning models typically generate a probability distribution over the known class labels, always selecting the most likely class. Ideally, for an unknown input, all classes should exhibit low probabilities, and setting a threshold based on uncertainty should reject these unknown classes. However, recent studies have shown that even inputs far from any known class can produce high probability/confidence scores \cite{bendale2016towards}. This leads to misleadingly high confidence scores even when the model’s predictions are incorrect. Neural networks, in particular, tend to produce overly confident predictions in such scenarios, creating a false sense of reliability \cite{lee2018simple}.

This paper addresses the challenges associated with concept drift in malware family detection by proposing a novel neural architecture and training paradigm tailored to this issue.  It is important to clarify that this work focuses solely on analyzing and differentiating between malware samples. Thus, we assume that all input samples are malware, and the goal is to identify and address variations within malware families rather than distinguishing between malicious and benign software. To achieve this, we leverage metric learning to map input samples to their respective families based on their proximity to the centroids of known classes. This method enhances the model's generalization ability by constructing a feature space that accurately reflects the similarities and differences between samples.
Additionally, we incorporate triplet loss to refine this feature space further, forming distinct, compact clusters that improve the accuracy of assignments. This approach ensures that samples from the same family are closely grouped, while those from different families are pushed apart. Importantly, our work specifically addresses concept drift in detecting new, unknown malware families, rather than focusing on the evolution of existing malware families. This distinction is critical because a related, yet underexplored, aspect of concept drift involves the automated detection of emerging malware families for multi-class classification purposes. While existing approaches may employ drift signaling techniques to determine when to retrain binary classification models, the automatic identification of new malware families —those that significantly deviate from historical data— poses a more intriguing challenge \cite{guerra2023android}.

Our experiments demonstrate the effectiveness of the proposed approach in detecting newly emerging malware families, offering valuable insights into maintaining robust defense mechanisms against rapidly evolving cyber threats. The results show a significant improvement in detection performance for new malware, providing a reliable and adaptive solution to keep pace with the dynamic landscape of cyber threats.

The main contributions of this paper are as follows:

\begin{itemize}
    \item We present a method that leverages metric learning and DBSCAN for family clustering to address concept drift in malware analysis.
    \item We propose a neural network architecture that utilizes an autoencoder and triplet loss for robust malware family detection.
    \item We extensively evaluate our approach using two relevant benchmark datasets for Android and Windows PE malware detection, as commonly used in the literature. 
\end{itemize}

The rest of the paper is organized as follows: \autoref{sec:related-work} covers related work, discussing existing approaches in malware detection, metric learning, and concept drift. \autoref{sec:method} explains our proposed method, detailing how we effectively utilize an autoencoder with triplet loss to differentiate between known and new malware families. \autoref{sec:dataset-and-setup} details the datasets we used and the experimental setup. \autoref{sec:evaluation} presents our experiments' results and evaluates our approach's effectiveness.  \autoref{sec:limitations} discusses the limitations of our study, highlighting potential areas for improvement. Finally, \autoref{sec:conclusion} concludes the paper and discusses potential future work.

\section{Related work} \label{sec:related-work}

\subsection{Machine Learning in Malware Detection}
Numerous studies have leveraged machine learning techniques to enhance the detection of malware and its various families by utilizing a range of features. These include static features, which are extracted from the malware without executing it to reveal the underlying code structure, such as Application Programming Interface (API) function calls \cite{ahmadi2016novel, sami2010malware}, bytes \cite{jain2011byte, fuyong2017malware} and opcodes \cite{mclaughlin2017deep, kakisim2022sequential, yuxin2019malware}. Additionally, some research has focused on using a set of these static features \cite{arp2014drebin, anderson2018ember, yang2021bodmas}. Dynamic features, on the other hand, capture the behavior of malware during its execution and include data, such as API call traces \cite{salehi2017maar, uppal2014malware}, instruction traces \cite{o2016detecting}, and network traffic \cite{boukhtouta2016network}. Some studies have furthered this approach by converting these inputs into visual formats, aiding in recognizing malicious patterns \cite{ahmadi2016novel, rezende2017malicious}. Machine learning thus plays a crucial role in strengthening cybersecurity measures against diverse threats.

\subsection{Metric Learning}
Recent advancements in metric learning have demonstrated their efficacy in learning high-quality data representations for tasks involving object differentiation and semantic similarity across various fields, such as computer vision \cite{hermans2017defense, dong2018triplet, mccartney2022zero}, audio processing \cite{oord2018representation, ghaleb2019metric, chung2020defence}, and bioinformatics \cite{xu2016multi, luo2019novel}. Metric learning has also garnered significant attention from security researchers due to its potential to enhance malware detection capabilities. \citeauthor{wu2022contrastive} \cite{wu2022contrastive} introduce IFDroid, a system that leverages contrastive learning to enhance the robustness and accuracy of Android malware family classification. IFDroid trains an encoder to extract features resilient to code obfuscation by converting function call graphs into images. Similarly, \citeauthor{jurevcek2021application} \cite{jurevcek2021application} employ Particle Swarm Optimization to optimize feature weights for a weighted Euclidean distance metric, improving the accuracy of k-Nearest Neighbors (k-NN) classification. Building on these concepts, \citeauthor{liu2022fewm} \cite{liu2022fewm} model the execution behavior of malware as heterogeneous graphs, capturing interactions between entities, such as APIs, processes, and files. This approach uses data augmentations and graph attention networks to generate robust positive and negative samples, enabling effective few-shot detection of malware variants without extensive labeled data. \citeauthor{andresini2021autoencoder} \cite{andresini2021autoencoder} present a method for network intrusion detection that combines autoencoders and triplet networks to improve predictive accuracy by addressing data imbalance and enhancing the separation of normal and malicious network traffic.

\subsection{Concept Drift}
In recent years, various approaches have been proposed to address concept drift in different domains, particularly in malware and intrusion detection systems. \citeauthor{singh2012tracking} \cite{singh2012tracking} explored concept drift in malware detection by introducing new tracking methods and examining different types of malware evolution. Building on this, \citeauthor{jordaney2017transcend} \cite{jordaney2017transcend} developed the TRANSCEND framework, which identifies concept drift in classification models through statistical metrics and a conformal evaluator to assess model credibility and confidence. The framework was further refined in TRANSCENDENT \cite{barbero2022transcending}, which formalized the theoretical foundation of conformal evaluation and introduced new evaluators to enhance robustness. Additionally, CADE \cite{yang2021cade} employed contrastive learning to detect outliers and explain concept drift by mapping data samples into a low-dimensional space. In the realm of intrusion detection, \citeauthor{andresini2021insomnia} \cite{andresini2021insomnia} proposed the INSOMNIA framework, which integrates incremental, active, and transfer learning to adapt to non-uniform data distribution over time, thus maintaining the efficacy of intrusion detection models through continuous updates and active learning strategies. Moreover, \citeauthor{zola2023temporal} \cite{zola2023temporal} conducted a temporal analysis of distribution shifts in malware classification, proposing a three-step forensic exploration approach to understand model failures caused by concept drift. These diverse methodologies highlight the evolving landscape of concept drift management in cybersecurity applications.

Many approaches do not adequately consider the issue of concept drift, which can lead to a decline in detection accuracy over time \cite{ahmadi2016novel, wu2022contrastive, kakisim2022sequential}. Additionally, some methods overlook the complexity of dealing with multiple clusters within a single malware family and fail to address the presence of outliers effectively \cite{jordaney2017transcend, yang2021cade}. These gaps highlight the need for more robust and adaptive solutions in malware family detection.

\section{Method} \label{sec:method}

\begin{figure}[ht]
    \centering
    \includegraphics[width=\linewidth]{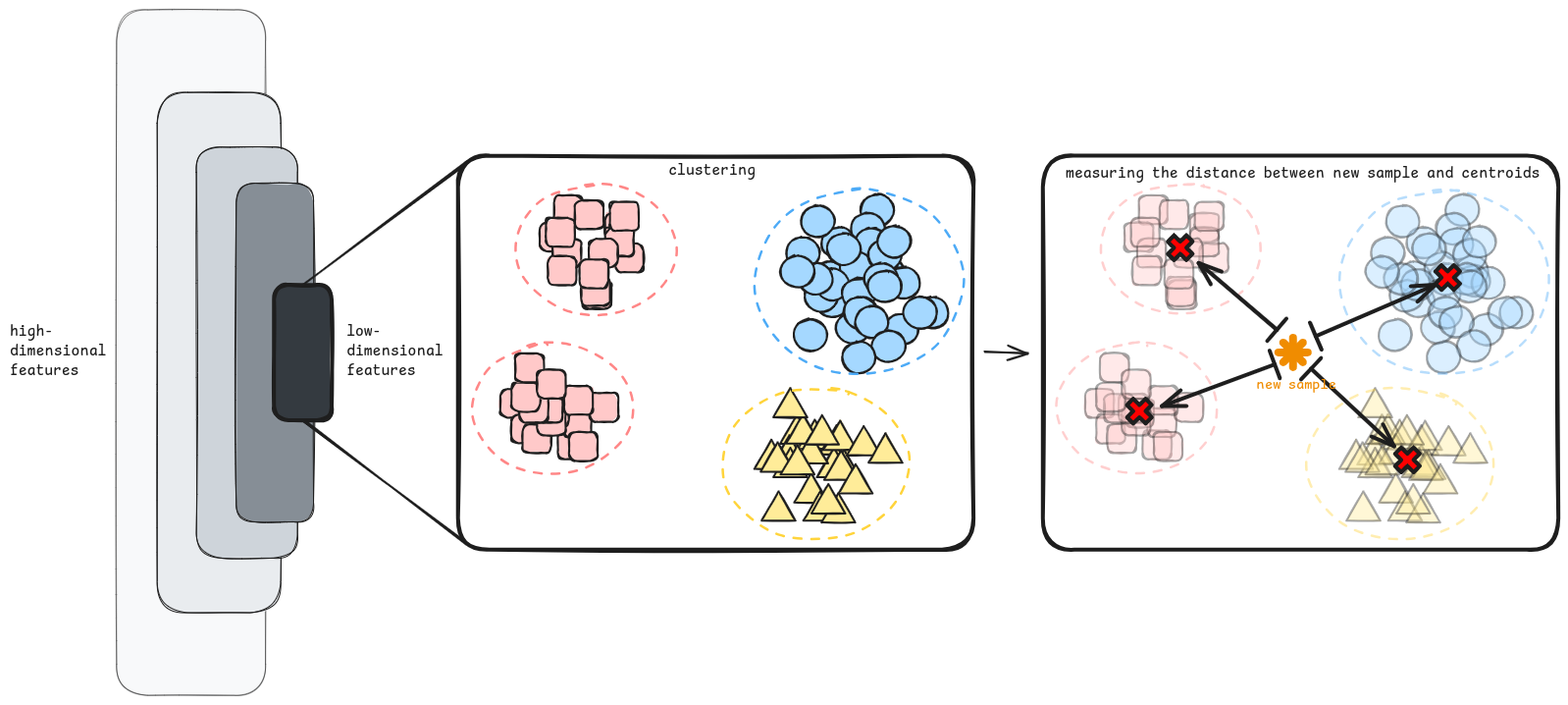}
    \caption{An overview of the method. }
    \label{fig:framework}
\end{figure}

Our method enhances the robustness and performance of detection models through a structured approach to address the challenges posed by concept drift in malware family classification. Initially, high-dimensional feature vectors are extracted from the malware samples. These vectors are then processed using an autoencoder, trained with triplet loss to reduce dimensionality while ensuring that similar points are closer together and dissimilar points are further apart in the latent space. Subsequently, clustering in the latent space using the DBSCAN algorithm is performed to identify sub-clusters within malware families and exclude outliers. Finally, centroids of these clusters are calculated to determine the classification of new samples based on their proximity to the closest centroid, compared against a pre-calculated threshold for cluster membership. The high-level process of the method is illustrated in Figure \ref{fig:framework}. This section outlines the approach and techniques we employed to maintain reliable classification performance in the face of the constantly evolving nature of malware.

We employ metric learning, a machine learning approach that focuses on defining a distance metric between data points to transform the data space. This method aims to bring similar points closer together while pushing dissimilar points further apart \cite{kaya2019deep}. In our implementation, we utilize triplet loss, which simultaneously considers pairs of similar and dissimilar points, enhancing the model’s capability to learn meaningful representations. Triplet loss operates by using triplets of data points: an anchor, a positive sample that is similar to the anchor, and a negative sample that is dissimilar \cite{hermans2017defense}. The objective is to ensure that the distance between the anchor and the positive sample is smaller than the distance between the anchor and the negative sample by at least a specified margin. This margin enforces a separation between similar and dissimilar pairs, driving the model to learn a more discriminative feature space. Additionally, working with high-dimensional feature vectors introduces the challenge of the curse of dimensionality. In high-dimensional spaces, distances between data points become less informative, complicating the task of distinguishing between different classes \cite{altman2018curse}. To overcome this, we incorporated an autoencoder to reduce the dimensionality of the data, thereby mitigating the effects of the curse of dimensionality and making distance measures more reliable. The autoencoder compresses the high-dimensional data into a lower-dimensional latent space, preserving the most critical features while discarding redundant information. This compression not only enhances the efficiency of distance computations but also helps in capturing the underlying structure of the data, making it easier to identify and distinguish between different classes.

After samples are projected into the latent space, samples from the same malware family are grouped, facilitating easier classification. However, since our goal is to analyze new malware families, a more fine-grained analysis is required. To achieve this, clustering in the latent space is used to group samples belonging to subgroups of the same malware families. This approach allows us to project new samples during testing and not only classify them as belonging to an existing or new family but also measure their deviation from existing families. Building on this foundation, we employed the DBSCAN \cite{schubert2017dbscan} clustering algorithm to identify multiple clusters within each class in the latent space obtained from the bottleneck of the autoencoder. Although we know the number of classes from the training data, DBSCAN helps us identify sub-clusters within these classes, which is crucial since a single class can contain multiple distinct clusters.

Ignoring these sub-clusters can result in misleadingly large distances between data points and their centroids. By employing DBSCAN, we can accurately detect these sub-clusters, leading to precise centroid calculations. This clustering process allows us to determine whether a new sample fits within existing classes or should be considered a new malware family or variant, pending expert verification. Additionally, DBSCAN effectively handles outliers, ensuring that anomalies do not distort the centroid calculations. 

To further refine our classification approach, we calculate the centroids of each cluster. Given that we use the DBSCAN algorithm, which does not include every point in a cluster and filters out outliers, we compute the centroids more accurately. The centroids are calculated by taking the mean of the data points within each cluster. This method benefits from DBSCAN's ability to handle outliers effectively, ensuring that these outliers do not distort the centroid calculations.

Once the clusters and their centroids are established, we determine whether a new sample belongs to an existing family by locating the closest centroid in the latent space and noting the family to which this centroid belongs. The distance between the new sample and the closest centroid is then calculated and compared with a pre-calculated threshold for the identified family. Each family's threshold is determined by the distance from the centroid to the furthest point within the cluster (excluding outliers). This threshold acts as a boundary to decide cluster membership. If the calculated distance for the new sample is less than or equal to the threshold, the sample is classified as belonging to that family. Otherwise, it is considered as not belonging to any existing family and may be flagged as a potential new family or variant.

The threshold for each family is calculated during the clustering process with DBSCAN. For each cluster, the threshold is defined as the distance from the centroid to the furthest point within the cluster. This is feasible because DBSCAN effectively excludes outliers, ensuring that the threshold represents the maximum distance within the core points of the cluster.

By employing DBSCAN and centroid calculation, our method can accurately determine whether a new sample fits within existing classes or should be considered a new malware family or variant, pending expert verification. This approach not only enhances detection performance but also provides a reliable mechanism to handle the dynamic nature of malware, maintaining the model’s robustness over time.

\subsection*{Impact of triplet loss}

The triplet loss function is a powerful technique used in machine learning to enhance the discriminative ability of models \cite{hermans2017defense}. It operates by optimizing the distance between samples in such a way that similar samples (belonging to the same class) are brought closer together, while dissimilar samples (belonging to different classes) are pushed farther apart. 

Given an anchor sample $x_a$, a positive sample $x_p$ of the same class, and a negative sample $x_n$ of a different class, the triplet loss function aims to ensure that the distance between the anchor and positive samples is less than the distance between the anchor and negative samples by at least a margin $\alpha$. The triplet loss function $L$ can be defined as:

\begin{equation}
    L(x_a, x_p, x_n) = \max \left\{ 0, D(x_a, x_p) - D(x_a, x_n) + \alpha \right\}
\end{equation}

where:

\begin{equation}
    D(x_i, x_j) = \| f(x_i) - f(x_j) \|^2
\end{equation}

represents the squared Euclidean distance between the embedding vectors of two samples, produced by the embedding function $f$. The margin $\alpha$ defines the desired separation between positive and negative pairs.

In practice, the loss is computed over a batch of triplets, an anchor input (input pair), a positive input (similar pair), and a negative input (dissimilar pair), and the objective is to minimize the total loss across the batch. In triplet loss, a margin enforces a distinct separation, where distances smaller than the margin do not contribute to the loss function. Assuming we have a set of triplets $\{(x_{a_i}, x_{p_i}, x_{n_i})\}_{i=1}^{N}$, the overall triplet loss can be defined as:

\begin{equation}
    L_{\text{batch}} = \frac{1}{N} \sum_{i=1}^{N} L(x_{a_i}, x_{p_i}, x_{n_i})  
\end{equation}

This formulation forces the network to learn an embedding space where positive pairs (anchor and positive samples) are closer than negative pairs (anchor and negative samples), with a margin $\alpha$ separating them.

\subsection*{Impact of DBSCAN}

Selecting an appropriate clustering algorithm is crucial for accurate identification and analysis. Popular clustering methods like K-means were not suitable for our needs due to several reasons. Firstly, K-means require the number of clusters to be specified beforehand, which is challenging in our scenario because the number of clusters is unknown and can vary significantly. A single class could potentially have multiple clusters due to intraclass variants. Secondly, K-means tries to include all data points in a cluster, which is not ideal for real-world data where outliers, such as incorrectly labeled data or genuine anomalies, are present. These outliers can distort the clustering results and reduce the overall accuracy. Additionally, K-means assumes that clusters are spherical and equally sized, which is rarely the case in malware datasets \cite{ahmed2020k}. Malware families can exhibit diverse and complex structures that do not fit the spherical assumption.

Given these limitations, we considered DBSCAN \cite{ester1996density} as a potential solution. DBSCAN is effective in identifying clusters of arbitrary shapes and sizes and is robust in detecting outliers. It operates on the principle that regions of high data density are separated by regions of low density. DBSCAN requires two parameters: epsilon $\varepsilon$, which defines the radius for neighborhood search, and minPts, the minimum number of points required to form a dense region. The challenge with DBSCAN lies in setting these parameters correctly, especially for datasets with varying densities. To address this, we employed several heuristics and techniques discussed in the literature. Firstly, we set the minPts parameter based on the dimensionality of the data. A common heuristic is to set minPts to twice the number of dimensions, i.e., $minPts = 2 \times dim$ \cite{schubert2017dbscan}. Choosing the appropriate value for $\varepsilon$ is more challenging. One effective method is to use the k-distance plot, where we plot the distance to the k-th nearest neighbor for each point in the dataset. A sharp bend in this plot, known as the "knee" or "elbow," often indicates a good choice for $\varepsilon$. In practice, the value of $\varepsilon$ should be chosen as small as possible to capture the most relevant clusters without merging distinct ones. 

In our case, there is another advantage of using DBSCAN. One crucial parameter to decide is the distance threshold metric, which helps us determine whether samples exceed or fall within the distance threshold. We utilized the DBSCAN algorithm to establish this distance threshold. The DBSCAN algorithm identifies clusters in the provided data, finding one or multiple clusters if they exist. Importantly, it does not force every point into a cluster; instead, it marks points that do not belong to any cluster as noise or outliers. This feature is advantageous for determining a robust threshold that accounts for outliers. We set the highest distance between a point and a centroid within a cluster as the threshold.

\begin{figure}[ht]
    \centering
    \includegraphics[width=\linewidth]{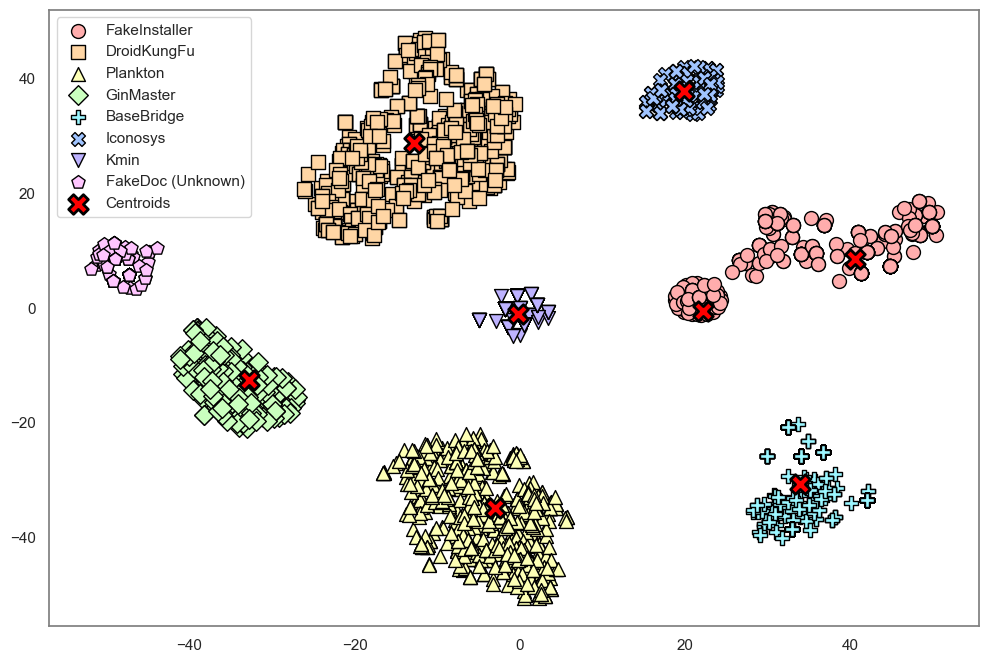}
    \caption{Clustering results using DBSCAN algorithm.}
    \label{fig:multiple_clusters}
\end{figure}

To demonstrate the effectiveness of DBSCAN, we present Figure \ref{fig:multiple_clusters} where multiple clusters within a single class are shown. Specifically, using DBSCAN, we detected two clusters in one of the classes, FakeInstaller. This approach decreased the mean distance between a sample and its centroid by 26\%, from 1.46 to 1.07. This significant reduction in mean distance highlights the effectiveness of DBSCAN in accurately separating clusters within a class, ultimately improving the clustering performance.

\begin{figure*}[ht]
    \centering
    \subfloat[Original feature space.]{\includegraphics[width=0.66\columnwidth]{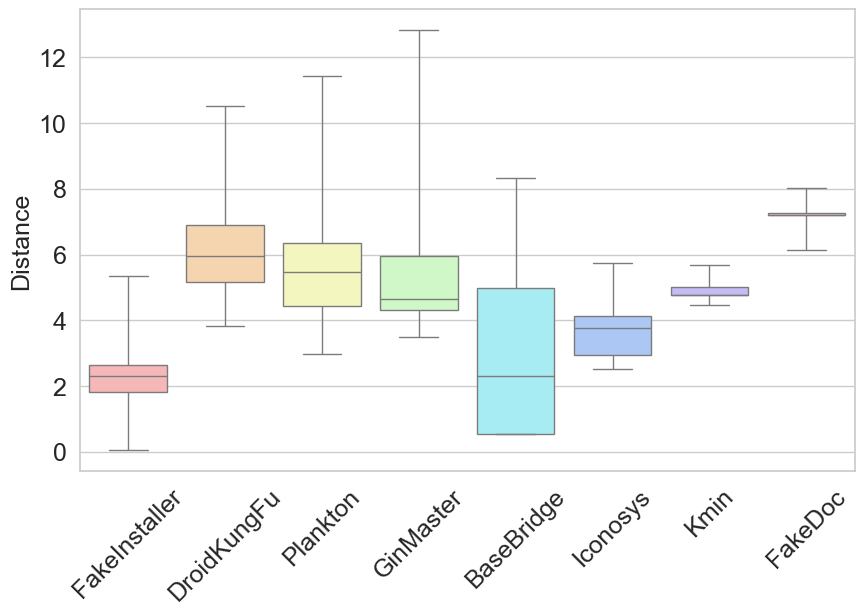}}
    \subfloat[Vanilla AE feature space.]{\includegraphics[width=0.66\columnwidth]{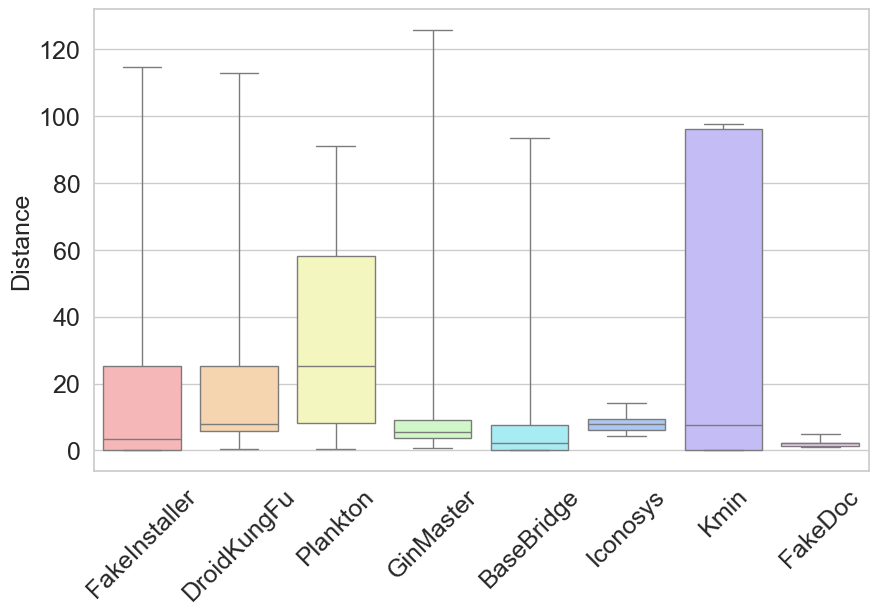}}
    \subfloat[Triplet autoencoder feature space.]{\includegraphics[width=0.66\columnwidth]{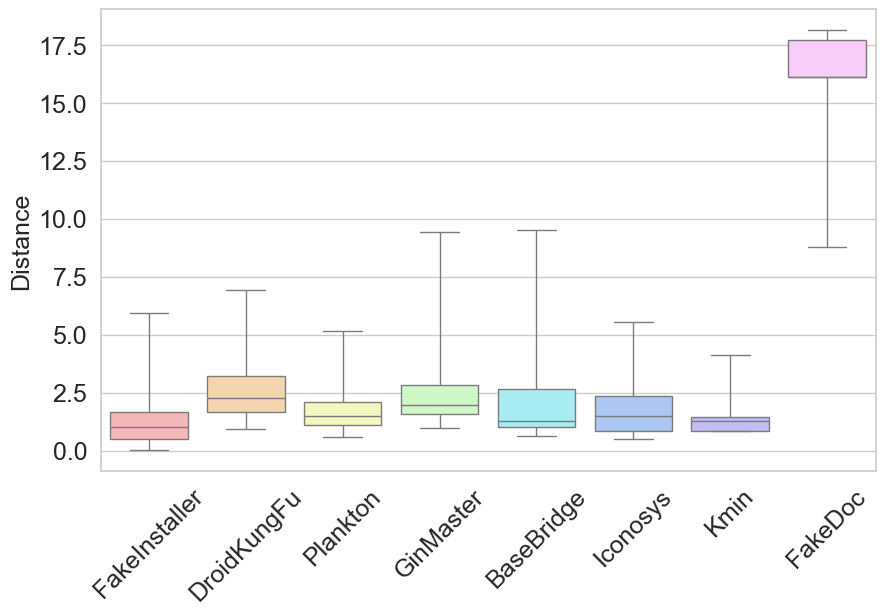}}
    \caption{Boxplot diagrams showing the distances between samples and their family centroids for three feature representations: original features, vanilla autoencoder features, and triplet autoencoder features.}
    \label{fig:boxplot}
\end{figure*}

\section{Dataset and Setup} \label{sec:dataset-and-setup}

\subsection{Drebin dataset}

We utilized the Drebin dataset, an Android malware dataset containing 5,560 malware samples and 123,453 benign applications compiled between August 2010 and October 2012 \cite{arp2014drebin}. Drebin extracts features from the Android Manifest file, such as hardware components, requested permissions, app components, and filtered intents, as well as features from the app's disassembled code, including restricted API calls, used permissions, suspicious API calls, and network addresses.

For our analysis, we selected families with a minimum of 100 malware samples, reducing the dataset to 8 families and a total of 3,317 samples (\autoref{tbl:drebin_sample_distribution}). We followed a previous study’s strategy of splitting the dataset into training and testing sets using an 80:20 ratio, based on the malware creation timestamps \cite{yang2021cade}. This temporal split ensures that our evaluation reflects real-world scenarios, where models encounter newer malware after being trained on older data, thus providing a robust evaluation against established benchmarks.

\begin{table}[ht]
    \caption{Sample Distribution of Drebin Dataset.}
    \begin{center}
        \begin{tabular}{cc}
            \toprule
            \textbf{Family} & \textbf{Number of Samples} \\
            \midrule
            FakeInstaller & 925           \\
            DroidKungFu   & 667           \\
            Plankton      & 625           \\
            GingerMaster  & 339           \\
            BaseBridge    & 330           \\
            Iconosys      & 152           \\
            Kmin          & 147           \\
            FakeDoc       & 132           \\
            \bottomrule
        \end{tabular}
    \end{center}
    \label{tbl:drebin_sample_distribution}
\end{table}

\subsection{BODMAS dataset}

We utilized the BODMAS dataset, originally consisting of 57,293 malware samples and 77,142 benign samples, for a total of 134,435 samples \cite{yang2021bodmas}. Each sample is represented by a 2,381-dimensional feature vector, extracted through static analysis. These feature vectors include elements such as general file information (file size, imported/exported functions, and section data like relocations, resources, and signatures), header information (machine type, subsystem, and image versions), imported and exported functions, and section information (section names, entropy, and virtual size). Additionally, byte histograms, byte entropy histograms, and string information (e.g., URLs, registry keys, and string entropy) are included to capture statistical properties. The dataset also records timestamp metadata, indicating when each sample was first seen on VirusTotal.

In our preprocessing, we performed several preprocessing steps to refine the subset used in our analysis. We excluded packed malware samples, as packers encrypt and compress the code, making it difficult to carry out accurate drift detection \cite{singh2012tracking, fernando2022fesa, zola2023temporal}. Additionally, we focused only on malware families with more than 1,500 samples, resulting in the inclusion of seven families for our study (\autoref{tbl:bodmas_sample_distrubution}).

We split the dataset into training and testing sets in an 80:20 ratio based on the first-seen timestamps to simulate real-world settings and evaluate the model's performance under more realistic conditions. The training data consists of samples from August 2019 to July 2020, while the testing data includes samples from July 2020 to September 2020. This temporal split helps minimize experimental bias and aligns with recommendations from previous work \cite{pendlebury2019tesseract}.

\begin{table}[ht]
    \caption{Sample Distribution of BODMAS Dataset.}
    \begin{center}
        \begin{tabular}{cc}
            \toprule
            \textbf{Family} & \textbf{Number of Samples} \\
            \midrule
            berbew &  1741   \\
            dinwod &  1942   \\
            ganelp &  1413   \\
            mira &  1526     \\
            sfone &  3218    \\
            sillyp2p & 3012 \\
            small &  3606    \\
            \bottomrule
        \end{tabular}
    \end{center}
    \label{tbl:bodmas_sample_distrubution}
\end{table}

\subsection{Experimental Setup}

To evaluate the robustness and accuracy of our proposed method, we designed a comprehensive experimental setup involving preprocessing, model training, and validation stages. This section outlines the procedures and configurations employed to ensure the credibility and reproducibility of our results. The experiments were conducted on two prominent malware datasets, Drebin and BODMAS, each requiring tailored preprocessing steps to handle their specific characteristics.

We began by implementing a variance threshold to filter out features with low variance in both datasets. Data was then split into training and testing sets based on timestamps, utilizing the malware creation time for Drebin and the first-seen date (based on VirusTotal) for BODMAS, ensuring a temporal allocation. The Drebin dataset was modeled using neural network layers consisting of 1376 input neurons, followed by 1024, 256, and 32 neurons. The BODMAS dataset's model architecture included layers with 2381, 1024, 256, and 32 neurons. For training, the vanilla autoencoder utilized mean squared error as its loss function, whereas the triplet autoencoder employed a combination of triplet loss and reconstruction loss. Triplets were selected by including one sample from a random class, one sample from the same class, and one from a different class.

To validate our model's performance, we used an 80:20 split for both datasets, with the training set comprising older samples and the testing set consisting of more recent samples. This temporal split simulates real-world scenarios where models are deployed and subsequently encounter new malware.

In our experiments, we simulate the presence of drifting samples by systematically excluding one malware family from the training data in each iteration. For example, if we exclude the FakeInstaller family, the remaining families are used to train the model, and during testing, the previously excluded family (e.g., FakeInstaller) is reintroduced alongside the test sets of the other families. This approach creates a scenario where the model encounters 'unknown' or 'drifting' families, allowing us to evaluate its ability to manage concept drift. This technique aligns with approaches used in previous studies \cite{jordaney2017transcend, yang2021cade}. Although this method effectively labels drifted samples (since they are excluded from training), it has a limitation: newer samples from known families in the test set might exhibit drift or initiate drifting, but without specific labels to denote this finer level of drift, they will not be classified as such.

\section{Evaluation} \label{sec:evaluation}

\subsection{Experiment on Android malware dataset (Drebin)}

In Figure \ref{fig:boxplot}, we present the distance boxplot diagrams of the distance between every sample and its family centroid for three different feature representations: the original features (1376-dimensional features), the vanilla autoencoder (32-dimensional features), and the triplet autoencoder (32-dimensional features). We observe that the distances from centroids are significantly smaller by comparing the original feature space to the triplet autoencoder feature space. This indicates that the samples of the same class are more tightly clustered and closer to each other in the triplet autoencoder feature space.

\begin{figure*}[ht]
    \centering
    \includegraphics[width=\linewidth, trim=10 10 10 10, clip]{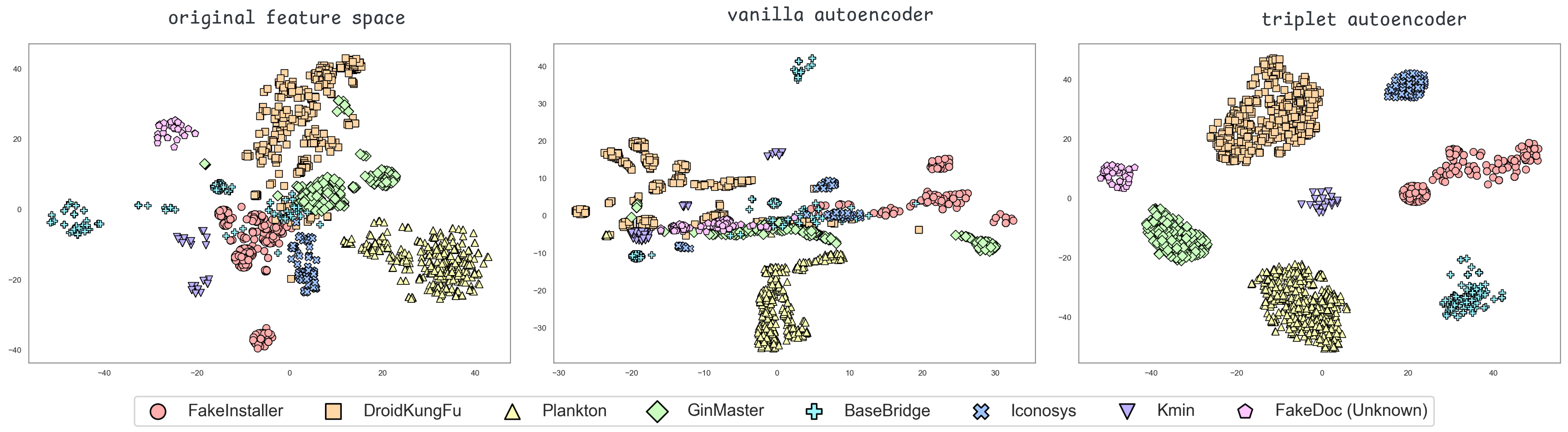}
    \caption{t-SNE diagrams of original features space, vanilla autoencoder and triplet autoencoder.}
    \label{fig:tsne}
\end{figure*}

When comparing the vanilla autoencoder to the triplet autoencoder, we also find that the distances in the triplet autoencoder are more compact. Additionally, we conducted an experiment where one class was designated as an unknown class by excluding any samples from that class during the training phase for both the vanilla autoencoder and the triplet autoencoder. This setup allows us to test the models' ability to handle out-of-distribution samples.

The results show that the distances between the unknown class samples and the closest centroid are much larger, indicating that the unknown class is effectively separated from the known classes. This separation suggests that the triplet autoencoder successfully differentiates between known and unknown classes, enhancing its robustness to data drift and unseen samples.

To further illustrate the separation of classes, we also provide t-distributed Stochastic Neighbor Embedding (t-SNE) graphs of the same data used to generate the boxplots in Figure \ref{fig:tsne}. These t-SNE visualizations offer a clearer picture of how the different feature representations perform in terms of clustering \cite{van2008visualizing}. By comparing the t-SNE plots of the vanilla autoencoder and the triplet autoencoder, it becomes evident that the clusters in the triplet autoencoder are more distinct and tightly packed. This demonstrates the superior capability of the triplet autoencoder in creating a well-defined and separated feature space.

Table \ref{tbl:fscore-drebin} provides comprehensive details about our experiments. The first two columns list the labels of the malware families, along with the family names that were excluded from the training set and reserved solely for testing. The table also presents the F1 scores, which measure the performance of the models in terms of both precision and recall. Three methods are compared in our evaluation, CADE \cite{yang2021cade}, triplet autoencoder with Median Absolute Deviation (MAD) threshold, and our proposed approach. MAD, calculated using the formula ${MAD} = {median}(|X_i - {median}(X)|)$, serves as a measure of statistical dispersion, and is used to determine the distance threshold. Including MAD allows us to compare the performance of the threshold using DBSCAN. Since the training model structure is the same for both the MAD and DBSCAN-based methods —both using the triplet autoencoder— the only difference lies in how the threshold is determined to decide whether each sample belongs to existing families. The results in the table demonstrate that the DBSCAN-based method consistently outperforms MAD method in terms of the overall F1 score. Even though in some cases MAD's performance is closer to that of DBSCAN, it is important to note that MAD requires a coefficient determined empirically. In contrast, DBSCAN calculates the threshold automatically without requiring a manual coefficient. When comparing CADE, which uses contrastive loss and MAD threshold, and our method, which employs triplet loss along with the DBSCAN threshold, our method demonstrates superior performance overall.

\begin{table*}[!ht]
    \caption{Performance of the models. The family column indicates which malware family was excluded from training and kept for testing to simulate drifting samples.}
    \begin{center}
        \begin{tabular}{ccccccc}
            \toprule
            \textbf{Family} & \makecell{\textbf{No. of} \\ \textbf{known samples}} & \makecell{\textbf{No. of} \\ \textbf{unknown samples}} & \makecell{\textbf{F1 score} \\ \textbf{CADE \cite{yang2021cade}}} & \makecell{\textbf{F1 score} \\ \textbf{MAD}} & \makecell{\textbf{F1 score} \\ \textbf{DBSCAN}} \\
            \midrule
                FakeInstaller    & 478 & 925 & 0.86 & 0.95 & \textbf{0.95}  \\
                DroidKungFu      & 529 & 667 & 0.87 & 0.89 & \textbf{0.90}  \\
                Plankton         & 538 & 625 & 0.77 & 0.90 & \textbf{0.87}  \\
                GinMaster        & 595 & 339 & 0.63 & 0.84 & \textbf{0.85}  \\
                BaseBridge       & 597 & 330 & 0.59 & 0.97 & \textbf{0.98}  \\
                Iconosys         & 632 & 152 & 0.42 & 0.46 & \textbf{0.65}  \\
                Kmin             & 633 & 147 & 0.40 & 0.63 & \textbf{0.62}  \\
                FakeDoc          & 636 & 132 & 0.38 & 0.56 & \textbf{0.66}  \\
            \midrule
            \textbf{Overall}     & \textbf{4638} & \textbf{3317} & \textbf{0.62} & \textbf{0.78} & \textbf{0.81} \\
            \bottomrule
        \end{tabular}
    \end{center}
    \label{tbl:fscore-drebin}
\end{table*}

\subsection{Experiment on Windows PE malware dataset (BODMAS)}
We also used the BODMAS dataset to measure the performance of our method. This dataset brings two different advantages: it is a more recent dataset, and it is a Windows PE dataset, representing a different operating system, which helps us measure the generalizability of our method.

The results, summarized in Table \ref{tbl:fscore-bodmas}, show that our method performs well on the Windows PE malware dataset. This performance demonstrates the robustness and generalizability of our method across different types of malware and operating systems. By using the BODMAS dataset, we validated that our approach is not limited to a specific type of malware or operating system but can be effectively applied to various scenarios. This enhances the overall reliability and applicability of our malware detection method, providing a practical and robust solution for cybersecurity challenges across different platforms.

\begin{table}[ht]
    \caption{Performance of the models. The family column indicates which malware family was excluded from training and kept for testing to simulate drifting samples.}
    \begin{center}
        \begin{tabular}{cccc}
            \toprule
            \textbf{Family} & \makecell{\textbf{No. of} \\ \textbf{known samples}} & \makecell{\textbf{No. of} \\ \textbf{unknown samples}} & \textbf{F1 score} \\
            \midrule
                berbew    & 2817 & 1741 & 0.99  \\
                dinwod    & 2634 & 1942 & 0.96  \\
                ganelp    & 2636 & 1413 & 0.97  \\
                mira      & 2470 & 1526 & 0.58  \\
                sfone     & 2231 & 3218 & 0.51  \\
                sillyp2p  & 2737 & 3012 & 0.83  \\
                small     & 1473 & 3606 & 0.96  \\
            \bottomrule
        \end{tabular}
    \end{center}
    \label{tbl:fscore-bodmas}
\end{table}

\section{Limitations} \label{sec:limitations}
While our study provides valuable insights, it is important to acknowledge several limitations. Our analysis considers only one family as the unknown family. However, in real-world scenarios, there can be multiple unknown families, which could affect the clustering results and the interpretation of the data. Moreover, although we automatically determined the parameters of the DBSCAN algorithm based on the method outlined in \cite{schubert2017dbscan}, parameters, such as $\varepsilon$ and the minimum number of points (minPts) could be further fine-tuned depending on the specific context and dataset characteristics. Fine-tuning these parameters might lead to more accurate and meaningful clustering results. Additionally, our study did not include any packed samples because the encryption and compression of code by packers make accurate drift detection challenging.


\section{Conclusion and Future Work} \label{sec:conclusion}
This paper presents an approach to addressing concept drift in malware family detection, specifically focusing on the emergence of new malware families. Our method effectively differentiates between known and new malware families by leveraging a triplet autoencoder and the DBSCAN clustering algorithm. We validated our method using two prominent datasets, Drebin and BODMAS, representing Android and Windows PE malware. The results demonstrate that our approach significantly improves the detection performance for new malware families. This robust and reliable solution addresses the dynamic nature of cyber threats, ensuring that detection models remain effective over time. Our contributions include applying metric learning and clustering techniques to improve malware family classification in the face of concept drift, providing a practical framework for ongoing cybersecurity efforts. 

Moving forward, we aim to explore retraining strategies to adapt our model to evolving malware behaviors more effectively. This will involve identifying the features and factors contributing to malware drift, allowing us to understand the dynamics of malware evolution better. By incorporating these insights, we aim to develop more robust detection mechanisms to maintain high performance even as malware tactics change.

\section*{Acknowledgments}
Numan Halit Guldemir is supported by the Republic of Türkiye Ministry of National Education (MoNE-1416/YLSY).
\printbibliography

@article{altman2018curse,
  title={The curse (s) of dimensionality},
  author={Altman, Naomi and Krzywinski, Martin},
  journal={Nat Methods},
  volume={15},
  number={6},
  pages={399--400},
  year={2018}
}

@article{lee2018simple,
  title={A simple unified framework for detecting out-of-distribution samples and adversarial attacks},
  author={Lee, Kimin and Lee, Kibok and Lee, Honglak and Shin, Jinwoo},
  journal={Advances in neural information processing systems},
  volume={31},
  year={2018}
}

@article{guerra2023android,
  title={Android malware detection: mission accomplished? A review of open challenges and future perspectives},
  author={Guerra-Manzanares, Alejandro},
  journal={Computers \& Security},
  pages={103654},
  year={2023},
  publisher={Elsevier}
}

@inproceedings{jordaney2017transcend,
  title={Transcend: Detecting concept drift in malware classification models},
  author={Jordaney, Roberto and Sharad, Kumar and Dash, Santanu K and Wang, Zhi and Papini, Davide and Nouretdinov, Ilia and Cavallaro, Lorenzo},
  booktitle={26th USENIX security symposium (USENIX security 17)},
  pages={625--642},
  year={2017}
}

@inproceedings{ester1996density,
  title={A density-based algorithm for discovering clusters in large spatial databases with noise},
  author={Ester, Martin and others},
  booktitle={KDD},
  volume={96},
  number={34},
  pages={226--231},
  year={1996}
}

@inproceedings{barbero2022transcending,
  title={Transcending transcend: Revisiting malware classification in the presence of concept drift},
  author={Barbero, Federico and Pendlebury, Feargus and Pierazzi, Fabio and Cavallaro, Lorenzo},
  booktitle={2022 IEEE Symposium on Security and Privacy (SP)},
  pages={805--823},
  year={2022},
  organization={IEEE}
}

@inproceedings{yang2021cade,
  title={CADE: Detecting and explaining concept drift samples for security applications},
  author={Yang, Limin and Guo, Wenbo and Hao, Qingying and Ciptadi, Arridhana and Ahmadzadeh, Ali and Xing, Xinyu and Wang, Gang},
  booktitle={30th USENIX Security Symposium (USENIX Security 21)},
  pages={2327--2344},
  year={2021}
}

@inproceedings{andresini2021insomnia,
  title={Insomnia: Towards concept-drift robustness in network intrusion detection},
  author={Andresini, Giuseppina and Pendlebury, Feargus and Pierazzi, Fabio and Loglisci, Corrado and Appice, Annalisa and Cavallaro, Lorenzo},
  booktitle={Proceedings of the 14th ACM workshop on artificial intelligence and security},
  pages={111--122},
  year={2021}
}

@article{andresini2021autoencoder,
  title={Autoencoder-based deep metric learning for network intrusion detection},
  author={Andresini, Giuseppina and Appice, Annalisa and Malerba, Donato},
  journal={Information Sciences},
  volume={569},
  pages={706--727},
  year={2021},
  publisher={Elsevier}
}

@article{schubert2017dbscan,
  title={DBSCAN revisited, revisited: why and how you should (still) use DBSCAN},
  author={Schubert, Erich and Sander, J{\"o}rg and Ester, Martin and Kriegel, Hans Peter and Xu, Xiaowei},
  journal={ACM Transactions on Database Systems (TODS)},
  volume={42},
  number={3},
  pages={1--21},
  year={2017},
  publisher={ACM New York, NY, USA}
}

@article{lu2018learning,
  title={Learning under concept drift: A review},
  author={Lu, Jie and Liu, Anjin and Dong, Fan and Gu, Feng and Gama, Joao and Zhang, Guangquan},
  journal={IEEE transactions on knowledge and data engineering},
  volume={31},
  number={12},
  pages={2346--2363},
  year={2018},
  publisher={IEEE}
}

@article{ahmed2020k,
  title={The k-means algorithm: A comprehensive survey and performance evaluation},
  author={Ahmed, Mohiuddin and Seraj, Raihan and Islam, Syed Mohammed Shamsul},
  journal={Electronics},
  volume={9},
  number={8},
  pages={1295},
  year={2020},
  publisher={MDPI}
}

@inproceedings{singh2012tracking,
  title={Tracking concept drift in malware families},
  author={Singh, Anshuman and Walenstein, Andrew and Lakhotia, Arun},
  booktitle={Proceedings of the 5th ACM workshop on Security and artificial intelligence},
  pages={81--92},
  year={2012}
}

@article{fernando2022fesa,
  title={FeSA: Feature selection architecture for ransomware detection under concept drift},
  author={Fernando, Damien Warren and Komninos, Nikos},
  journal={Computers \& Security},
  volume={116},
  pages={102659},
  year={2022},
  publisher={Elsevier}
}

@article{hermans2017defense,
  title={In defense of the triplet loss for person re-identification},
  author={Hermans, Alexander and Beyer, Lucas and Leibe, Bastian},
  journal={arXiv preprint arXiv:1703.07737},
  year={2017}
}

@inproceedings{dong2018triplet,
  title={Triplet loss in siamese network for object tracking},
  author={Dong, Xingping and Shen, Jianbing},
  booktitle={Proceedings of the European conference on computer vision (ECCV)},
  pages={459--474},
  year={2018}
}

@article{mccartney2022zero,
  title={A zero-shot deep metric learning approach to brain--computer interfaces for image retrieval},
  author={McCartney, Ben and Devereux, Barry and Martinez-del-Rincon, Jesus},
  journal={Knowledge-Based Systems},
  volume={246},
  pages={108556},
  year={2022},
  publisher={Elsevier}
}

@article{oord2018representation,
  title={Representation learning with contrastive predictive coding},
  author={Oord, Aaron van den and Li, Yazhe and Vinyals, Oriol},
  journal={arXiv preprint arXiv:1807.03748},
  year={2018}
}

@article{ghaleb2019metric,
  title={Metric learning-based multimodal audio-visual emotion recognition},
  author={Ghaleb, Esam and Popa, Mirela and Asteriadis, Stylianos},
  journal={IEEE Multimedia},
  volume={27},
  number={1},
  pages={37--48},
  year={2019},
  publisher={IEEE}
}

@article{chung2020defence,
  title={In defence of metric learning for speaker recognition},
  author={Chung, Joon Son and Huh, Jaesung and Mun, Seongkyu and Lee, Minjae and Heo, Hee Soo and Choe, Soyeon and Ham, Chiheon and Jung, Sunghwan and Lee, Bong-Jin and Han, Icksang},
  journal={arXiv preprint arXiv:2003.11982},
  year={2020}
}

@article{xu2016multi,
  title={Multi-instance multi-label distance metric learning for genome-wide protein function prediction},
  author={Xu, Yonghui and Min, Huaqing and Song, Hengjie and Wu, Qingyao},
  journal={Computational biology and chemistry},
  volume={63},
  pages={30--40},
  year={2016},
  publisher={Elsevier}
}

@article{luo2019novel,
  title={A novel drug repositioning approach based on collaborative metric learning},
  author={Luo, Huimin and Wang, Jianxin and Yan, Cheng and Li, Min and Wu, Fang-Xiang and Pan, Yi},
  journal={IEEE/ACM transactions on computational biology and bioinformatics},
  volume={18},
  number={2},
  pages={463--471},
  year={2019},
  publisher={IEEE}
}

@article{wu2022contrastive,
  title={Contrastive learning for robust android malware familial classification},
  author={Wu, Yueming and Dou, Shihan and Zou, Deqing and Yang, Wei and Qiang, Weizhong and Jin, Hai},
  journal={IEEE Transactions on Dependable and Secure Computing},
  year={2022},
  publisher={IEEE}
}

@article{jurevcek2021application,
  title={Application of distance metric learning to automated malware detection},
  author={Jure{\v{c}}ek, Martin and L{\'o}rencz, R{\'o}bert},
  journal={IEEE Access},
  volume={9},
  pages={96151--96165},
  year={2021},
  publisher={IEEE}
}

@article{liu2022fewm,
  title={FewM-HGCL: Few-shot malware variants detection via heterogeneous graph contrastive learning},
  author={Liu, Chen and Li, Bo and Zhao, Jun and Zhen, Ziyang and Liu, Xudong and Zhang, Qunshi},
  journal={IEEE Transactions on Dependable and Secure Computing},
  year={2022},
  publisher={IEEE}
}

@inproceedings{ahmadi2016novel,
  title={Novel feature extraction, selection and fusion for effective malware family classification},
  author={Ahmadi, Mansour and Ulyanov, Dmitry and Semenov, Stanislav and Trofimov, Mikhail and Giacinto, Giorgio},
  booktitle={Proceedings of the sixth ACM conference on data and application security and privacy},
  pages={183--194},
  year={2016}
}

@inproceedings{sami2010malware,
  title={Malware detection based on mining API calls},
  author={Sami, Ashkan and Yadegari, Babak and Rahimi, Hossein and Peiravian, Naser and Hashemi, Sattar and Hamze, Ali},
  booktitle={Proceedings of the 2010 ACM symposium on applied computing},
  pages={1020--1025},
  year={2010}
}

@inproceedings{jain2011byte,
  title={Byte level n--gram analysis for malware detection},
  author={Jain, Sachin and Meena, Yogesh Kumar},
  booktitle={Computer Networks and Intelligent Computing: 5th International Conference on Information Processing, ICIP 2011, Bangalore, India, August 5-7, 2011. Proceedings},
  pages={51--59},
  year={2011},
  organization={Springer}
}

@inproceedings{fuyong2017malware,
  title={Malware detection and classification based on n-grams attribute similarity},
  author={Fuyong, Zhang and Tiezhu, Zhao},
  booktitle={2017 IEEE international conference on computational science and engineering (CSE) and IEEE international conference on embedded and ubiquitous computing (EUC)},
  volume={1},
  pages={793--796},
  year={2017},
  organization={IEEE}
}

@inproceedings{mclaughlin2017deep,
  title={Deep android malware detection},
  author={McLaughlin, Niall and Martinez del Rincon, Jesus and Kang, BooJoong and Yerima, Suleiman and Miller, Paul and Sezer, Sakir and Safaei, Yeganeh and Trickel, Erik and Zhao, Ziming and Doup{\'e}, Adam and others},
  booktitle={Proceedings of the seventh ACM on conference on data and application security and privacy},
  pages={301--308},
  year={2017}
}

@article{kakisim2022sequential,
  title={Sequential opcode embedding-based malware detection method},
  author={Kakisim, Arzu Gorgulu and Gulmez, Sibel and Sogukpinar, Ibrahim},
  journal={Computers \& Electrical Engineering},
  volume={98},
  pages={107703},
  year={2022},
  publisher={Elsevier}
}

@article{yuxin2019malware,
  title={Malware detection based on deep learning algorithm},
  author={Yuxin, Ding and Siyi, Zhu},
  journal={Neural Computing and Applications},
  volume={31},
  pages={461--472},
  year={2019},
  publisher={Springer}
}

@article{o2016detecting,
  title={Detecting obfuscated malware using reduced opcode set and optimised runtime trace},
  author={O’kane, Philip and Sezer, Sakir and McLaughlin, Kieran},
  journal={Security Informatics},
  volume={5},
  pages={1--12},
  year={2016},
  publisher={Springer}
}

@article{salehi2017maar,
  title={MAAR: Robust features to detect malicious activity based on API calls, their arguments and return values},
  author={Salehi, Zahra and Sami, Ashkan and Ghiasi, Mahboobe},
  journal={Engineering Applications of Artificial Intelligence},
  volume={59},
  pages={93--102},
  year={2017},
  publisher={Elsevier}
}

@inproceedings{uppal2014malware,
  title={Malware detection and classification based on extraction of API sequences},
  author={Uppal, Dolly and Sinha, Rakhi and Mehra, Vishakha and Jain, Vinesh},
  booktitle={2014 International conference on advances in computing, communications and informatics (ICACCI)},
  pages={2337--2342},
  year={2014},
  organization={IEEE}
}

@article{boukhtouta2016network,
  title={Network malware classification comparison using DPI and flow packet headers},
  author={Boukhtouta, Amine and Mokhov, Serguei A and Lakhdari, Nour-Eddine and Debbabi, Mourad and Paquet, Joey},
  journal={Journal of Computer Virology and Hacking Techniques},
  volume={12},
  pages={69--100},
  year={2016},
  publisher={Springer}
}

@inproceedings{rezende2017malicious,
  title={Malicious software classification using transfer learning of resnet-50 deep neural network},
  author={Rezende, Edmar and Ruppert, Guilherme and Carvalho, Tiago and Ramos, Fabio and De Geus, Paulo},
  booktitle={2017 16th IEEE international conference on machine learning and applications (ICMLA)},
  pages={1011--1014},
  year={2017},
  organization={IEEE}
}

@inproceedings{arp2014drebin,
  title={Drebin: Effective and explainable detection of android malware in your pocket.},
  author={Arp, Daniel and Spreitzenbarth, Michael and Hubner, Malte and Gascon, Hugo and Rieck, Konrad and Siemens, CERT},
  booktitle={Ndss},
  volume={14},
  pages={23--26},
  year={2014}
}

@article{anderson2018ember,
  title={Ember: an open dataset for training static pe malware machine learning models},
  author={Anderson, Hyrum S and Roth, Phil},
  journal={arXiv preprint arXiv:1804.04637},
  year={2018}
}

@inproceedings{yang2021bodmas,
  title={BODMAS: An open dataset for learning based temporal analysis of PE malware},
  author={Yang, Limin and Ciptadi, Arridhana and Laziuk, Ihar and Ahmadzadeh, Ali and Wang, Gang},
  booktitle={2021 IEEE Security and Privacy Workshops (SPW)},
  pages={78--84},
  year={2021},
  organization={IEEE}
}

@inproceedings{bendale2016towards,
  title={Towards open set deep networks},
  author={Bendale, Abhijit and Boult, Terrance E},
  booktitle={Proceedings of the IEEE conference on computer vision and pattern recognition},
  pages={1563--1572},
  year={2016}
}

@inproceedings{zola2023temporal,
  title={Temporal analysis of distribution shifts in malware classification for digital forensics},
  author={Zola, Francesco and Bruse, Jan Lukas and Galar, Mikel},
  booktitle={2023 IEEE European Symposium on Security and Privacy Workshops (EuroS\&PW)},
  pages={439--450},
  year={2023},
  organization={IEEE}
}

@inproceedings{pendlebury2019tesseract,
  title={TESSERACT: Eliminating experimental bias in malware classification across space and time},
  author={Pendlebury, Feargus and Pierazzi, Fabio and Jordaney, Roberto and Kinder, Johannes and Cavallaro, Lorenzo},
  booktitle={28th USENIX security symposium (USENIX Security 19)},
  pages={729--746},
  year={2019}
}

@online{AVTest,
    title = {Malware Statistics \& Trends Report},
    author = "AV-Test",
    year = {2023},
    url = {https://www.av-test.org/en/statistics/malware/},
    note = {Accessed: 14 July 2024}
}

@article{kaya2019deep,
  title={Deep metric learning: A survey},
  author={Kaya, Mahmut and Bilge, Hasan {\c{S}}akir},
  journal={Symmetry},
  volume={11},
  number={9},
  pages={1066},
  year={2019},
  publisher={MDPI}
}

@article{van2008visualizing,
  title={Visualizing data using t-SNE.},
  author={Van der Maaten, Laurens and Hinton, Geoffrey},
  journal={Journal of machine learning research},
  volume={9},
  number={11},
  year={2008}
}
\end{document}